\documentclass[conference]{IEEEtran}
\IEEEoverridecommandlockouts

\usepackage{cite}
\usepackage{amsmath,amssymb,amsfonts}
\usepackage{graphicx}
\usepackage{booktabs}
\usepackage{tabularx}
\usepackage{textcomp}
\usepackage{xcolor}
\usepackage{siunitx}
\usepackage{placeins}
\graphicspath{{figure_work/generated/}{figures/}}
\def\BibTeX{{\rm B\kern-.05em{\sc i\kern-.025em b}\kern-.08em
    T\kern-.1667em\lower.7ex\hbox{E}\kern-.125emX}}

\begin{document}

\title{Optimization Design and Simulation Validation of a Variable Stiffness Actuator Based on a Crossed Four-Bar Mechanism}

\author{
\begin{tabular*}{0.96\textwidth}{@{\extracolsep{\fill}}ccc@{}}
\begin{tabular}{c}
1\textsuperscript{st} Yuanlong Ji\\
\textit{School of Biological Science}\\
\textit{and Medical Engineering}\\
\textit{Beihang University}\\
Beijing, China
\end{tabular}
&
\begin{tabular}{c}
2\textsuperscript{nd} Ruizhe Jiang\\
\textit{School of Biological Science}\\
\textit{and Medical Engineering}\\
\textit{Beihang University}\\
Beijing, China
\end{tabular}
&
\begin{tabular}{c}
3\textsuperscript{rd} Xiangyu Xie\\
\textit{School of Mechanical Engineering}\\
\textit{and Automation}\\
\textit{Beihang University}\\
Beijing, China
\end{tabular}
\end{tabular*}\\[1.2ex]
\begin{tabular*}{0.96\textwidth}{@{\extracolsep{\fill}}ccc@{}}
\begin{tabular}{c}
4\textsuperscript{th} Junheng Lin\\
\textit{School of Mechanical Engineering}\\
\textit{and Automation}\\
\textit{Beihang University}\\
Beijing, China
\end{tabular}
&
\begin{tabular}{c}
5\textsuperscript{th} Dongrun Jin\\
\textit{School of Biological Science}\\
\textit{and Medical Engineering}\\
\textit{Beihang University}\\
Beijing, China
\end{tabular}
&
\begin{tabular}{c}
6\textsuperscript{th} Xingbang Yang*\\
\textit{School of Biological Science}\\
\textit{and Medical Engineering}\\
\textit{Beihang University}\\
Beijing, China\\
yangxingbang@buaa.edu.cn\\
*Corresponding author
\end{tabular}
\end{tabular*}
}

\maketitle

\begin{abstract}
This paper presents a bio-inspired antagonistic variable stiffness actuator (VSA) based on two crossed four-bar compliant transmission elastic units (CFB-CTEs). The design addresses the difficulty of combining nonlinear elastic shaping with low structural inertia in antagonistic VSA mechanisms. Inspired by the crossed constraint behavior of the anterior and posterior cruciate ligaments during knee flexion, the proposed actuator uses geometric transmission, elastic energy storage, and bilateral antagonistic arrangement to shape the output torque and equivalent stiffness. A multi-objective optimization model is established to balance torque tracking accuracy, equivalent inertia, and mass. The selected compromise design achieved a torque root-mean-square error (RMSE) of $0.883~\mathrm{N\,mm}$ and a total mass of $50.2~\mathrm{g}$. Its average equivalent inertia was reduced by about 78\% compared with an independent torque-only optimized design. An ADAMS multibody model was further built to verify the structural response under single-input, opposite-input, and same-input conditions. The results support the feasibility of the proposed crossed four-bar elastic unit as a lightweight nonlinear elastic branch for antagonistic VSAs.
\end{abstract}

\begin{IEEEkeywords}
variable stiffness actuator, crossed four-bar mechanism, compliant transmission, optimization design, multibody dynamics simulation
\end{IEEEkeywords}

\section{Introduction}
Variable stiffness actuators (VSAs) are important drive solutions for robotic joints that must balance interaction safety, mechanical compliance, and motion performance. In service robots, rehabilitation devices, and domestic assistance systems, joint actuators should provide low stiffness during contact to reduce impact risk, while maintaining sufficient stiffness during load bearing and positioning tasks \cite{desantis2008,haddadin2009}. Compared with series elastic actuators with fixed compliance, VSAs can physically regulate joint stiffness and therefore provide an adjustable compromise between safety and output performance \cite{bicchi2004,pratt1995,vanderborght2013,wolf2016,grioli2015}.

According to the arrangement of the drive and compliance regulation units, VSAs are commonly divided into antagonistic and independent architectures. Antagonistic VSAs use two opposing elastic branches acting on the same output joint. Their equilibrium position and stiffness are jointly determined by the coordinated or differential motion of the two inputs, which resembles biological muscle co-contraction \cite{laurin1991,english1999}. Representative antagonistic systems include VSA-II and bidirectional antagonistic variable stiffness joints \cite{tonietti2005,petit2010}. Other designs, such as MACCEPA, AwAS/AwAS-II, CompAct-VSA, and vsaUT-II, have advanced VSA design through adjustable preload, variable lever arms, compact elastic modules, or integrated mechanisms \cite{vanham2007,jafari2010,tsagarakis2011,jafari2011,groothuis2014}. These studies demonstrate the value of adjustable stiffness, but they also reveal structural challenges such as complex nonlinear transmission, multiple moving components, increased inertia, and difficulty in explicit modeling.

For a structurally novel VSA, analytical curve fitting alone is insufficient to demonstrate mechanical feasibility. Analytical models reveal how geometric parameters, elastic elements, and output stiffness are related, whereas multibody dynamics models can verify the structural response under realistic motion constraints and joint connections. Previous studies have used analytical, numerical, or multibody models to evaluate VSA dynamics, stiffness regulation, and mechanical response \cite{albu2010,hussain2018,wang2025}. However, bio-inspired nonlinear elastic branches still require validation at both the target-curve fitting level and the multibody reproduction level.

This paper proposes a bio-inspired antagonistic VSA composed of two crossed four-bar compliant transmission elastic units (CFB-CTEs). The main contributions are as follows. First, an antagonistic VSA architecture based on bilateral CFB-CTEs is proposed to convert crossed biological constraints into a manufacturable elastic shaping mechanism. Second, a structural optimization model is established by considering torque-curve fitting, inertia-related structural cost, and mass. The selected compromise design achieved a torque RMSE of $0.883~\mathrm{N\,mm}$ and a total mass of $50.2~\mathrm{g}$. Third, an ADAMS multibody model is developed to quantitatively verify the structural response under single-input, opposite-input, and same-input conditions. The corresponding joint-angle RMSE values were $8.69\times10^{-4}~\mathrm{rad}$, $6.58\times10^{-6}~\mathrm{rad}$, and $3.70\times10^{-3}~\mathrm{rad}$, respectively.

\section{Structural Design and Mechanical Modeling}
\subsection{Overall Architecture}
\begin{figure}[htbp]
\centering
\includegraphics[width=0.96\columnwidth]{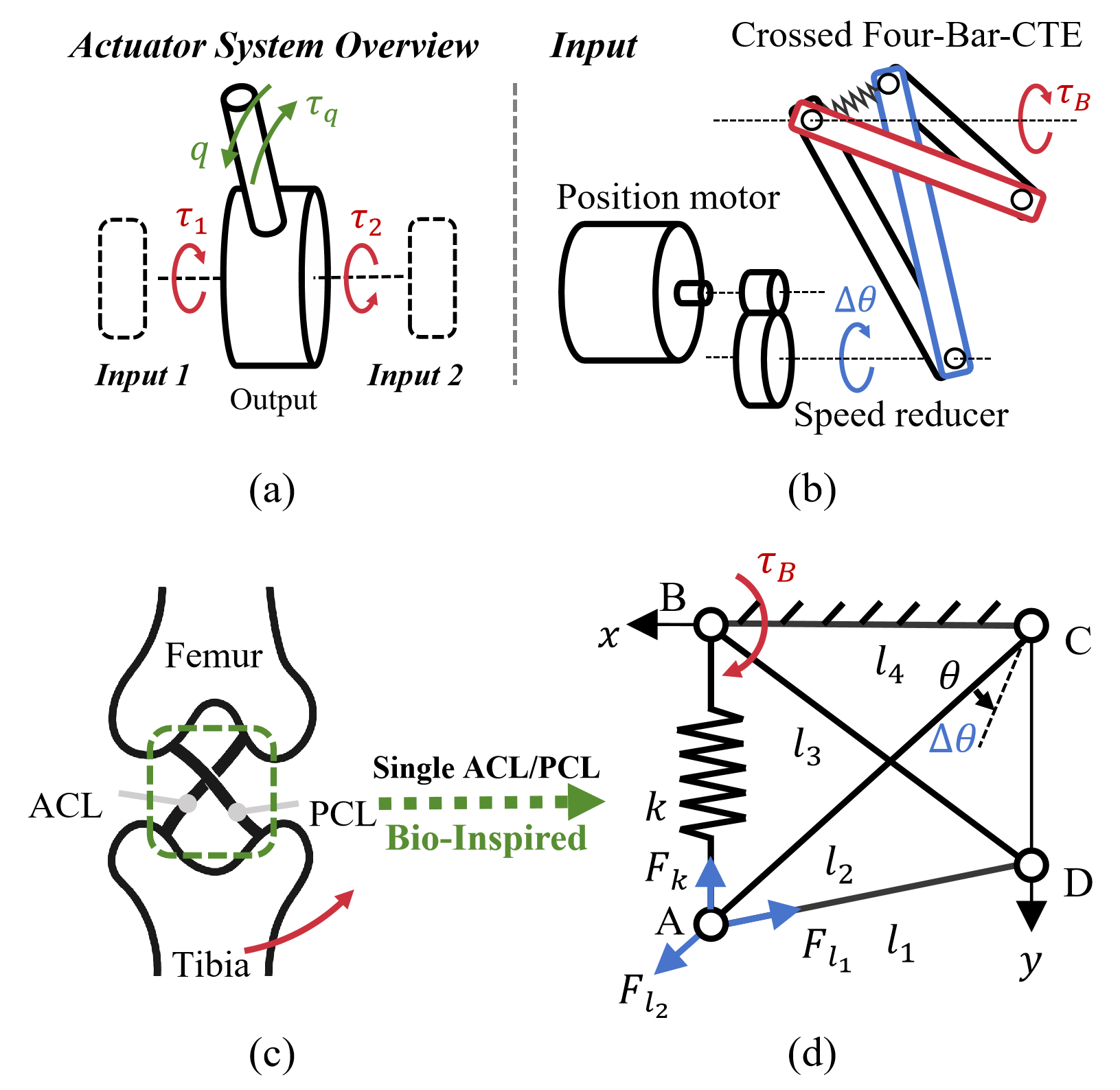}
\caption{Architecture, bio-inspired source, and mechanical model of the antagonistic VSA. (a) Overall dual-input antagonistic VSA architecture. (b) Transmission connection among a single-side input motor, reducer, and CFB-CTE. (c) Bio-inspired abstraction from ACL/PCL crossed constraints to a single-side CFB-CTE. (d) Geometric variables, spring force, and input torque relation of the CFB-CTE.}
\label{fig:bio}
\end{figure}

As shown in Fig.~\ref{fig:bio}(a) and Fig.~\ref{fig:bio}(b), the actuator consists of two input motors, two reduction transmissions, two mirrored CFB-CTEs, and one output joint. Input 1 and Input 2 act on the two elastic branches, respectively, and the output joint angle is denoted as $q$. The common-mode component of the two inputs drives the output position, whereas the relative pre-deformation component changes the equivalent output stiffness. Thus, stiffness regulation is achieved by antagonistic preload between the two elastic branches, rather than by changing the stiffness of a single spring.

Let the elastic torques generated by the two inputs be $\tau_1$ and $\tau_2$, and let the relative pre-deformation angle be $\Delta\theta$. The resultant output torque can be written as
\begin{equation}
\tau_q(q,\Delta\theta)=\tau_1(q+\Delta\theta)+\tau_2(q-\Delta\theta).
\label{eq:antagonistic_torque}
\end{equation}
Here, $\tau_1$ and $\tau_2$ are signed generalized torques projected onto the positive output coordinate. Therefore, \eqref{eq:antagonistic_torque} represents algebraic summation rather than the addition of two unsigned torque magnitudes.
The corresponding equivalent output stiffness is
\begin{equation}
K_q(q,\Delta\theta)=\frac{\partial \tau_q(q,\Delta\theta)}{\partial q}.
\label{eq:antagonistic_stiffness}
\end{equation}
Equations \eqref{eq:antagonistic_torque} and \eqref{eq:antagonistic_stiffness} show that both position output and stiffness regulation depend on the two elastic branches. When the relative pre-deformation changes, the equilibrium position can remain approximately unchanged, while the equivalent stiffness varies.

The structural design of an antagonistic VSA is not merely the selection of an elastic element. It must simultaneously achieve a desired torque profile, stiffness variation, and dual-input regulation within a limited workspace. A simple linear spring provides a linear force-displacement relation, and the output mainly depends on the spring stiffness and installation position. It is therefore difficult to satisfy torque magnitude, stiffness slope, and dimensional constraints over a broad angular range. This challenge is more pronounced in a bio-inspired antagonistic actuator, because the two elastic branches must generate output torque and also form adjustable stiffness through relative pre-deformation.

\subsection{Bio-Inspired CFB-CTE Design}
The anterior cruciate ligament (ACL) and posterior cruciate ligament (PCL) of the human knee exhibit crossed constraints and flexion-angle-dependent loading during knee flexion, as illustrated in Fig.~\ref{fig:bio}(c). The biomechanical basis is supported by numerical-experimental multi-bundle ligament models and in-situ force measurements of the ACL/PCL during flexion \cite{mommersteeg1996,li2004,carlin1996,fox1998}. This behavior differs from that of a simple linear spring, because the elastic response depends on the force direction, geometric constraint, and equivalent moment arm. The proposed engineering design extracts the crossed constraint, elastic energy storage, and nonlinear moment-arm features, without replicating the material or morphology of ligament tissues.

Based on this bio-inspired principle and the actuator architecture, a crossed four-bar compliant transmission elastic unit is designed. Its single-side input connection is shown in Fig.~\ref{fig:bio}(b). The unit consists of a crossed four-bar mechanism, a linear elastic element, and mounting components. The four-bar mechanism establishes a nonlinear geometric mapping between input angle, spring length, and moment arm. The elastic element stores energy, and the output side converts the spring force into torque for joint actuation.

Unlike a directly mounted torsional spring, the CFB-CTE changes both the spring elongation and the equivalent moment arm through mechanism posture. This allows a linear spring to generate a nonlinear output torque. The key feature is therefore not the spring material, but the reshaping of the elastic action path by the crossed four-bar geometry.

In the overall VSA, a single CFB-CTE only provides the torque-angle mapping of one elastic branch. To coordinate position output and stiffness regulation, two CFB-CTEs are mirrored on both sides of the output joint to form a bio-inspired antagonistic structure. The common-mode input mainly changes the resultant joint torque, whereas the opposite-mode input mainly changes the pre-deformation state of the two elastic branches. This arrangement provides a mechanical basis for approximate decoupling between position and stiffness, which is favorable for subsequent control design.

\subsection{Geometric and Torque Model}
The output characteristics of the CFB-CTE are determined by the link dimensions, installation angle, and spring stiffness. The design parameter vector is defined as
\begin{equation}
\boldsymbol p=[l_1,l_2,l_3,l_4,\theta_0,k]^T ,
\label{eq:p}
\end{equation}
where $l_1$--$l_4$ are the link lengths, $\theta_0$ is the initial installation angle, and $k$ is the spring stiffness. The mechanism input angle is denoted as $\theta$, and the relative angle change from the initial state is $\Delta\theta=\theta-\theta_0$. As the input angle changes, the positions of the two spring endpoints vary, producing spring elongation and output moment-arm variation.

\noindent\hspace*{1em}1) Geometric variables and spring elongation. Let the two spring endpoints be $A$ and $B$, with position vectors $\boldsymbol r_A(\theta)$ and $\boldsymbol r_B(\theta)$. The spring length is
\begin{equation}
l_s(\theta)=
\left\|
\boldsymbol r_A(\theta)-\boldsymbol r_B(\theta)
\right\|.
\label{eq:spring_length}
\end{equation}
The double vertical bars in \eqref{eq:spring_length} denote the Euclidean norm. The spring elongation is
\begin{equation}
\Delta l_s(\theta)=l_s(\theta)-l_0 ,
\label{eq:delta_l}
\end{equation}
where $l_0$ is the initial spring length. The corresponding spring force is
\begin{equation}
F_s(\theta)=k\Delta l_s(\theta).
\label{eq:spring_force}
\end{equation}
Equations \eqref{eq:spring_length}--\eqref{eq:spring_force} indicate that although the spring is linear, its elongation is governed by the mechanism geometry. The nonlinear structural output mainly results from the geometric shaping of the spring deformation path.

\noindent\hspace*{1em}2) Output torque and equivalent moment arm. As shown in Fig.~\ref{fig:bio}(d), the output torque of the CFB-CTE is jointly determined by the spring force and equivalent moment arm. If the equivalent moment arm of the spring force with respect to the output axis is $d_e(\theta)$, the output torque is
\begin{equation}
\tau_B(\theta)=F_s(\theta)d_e(\theta).
\label{eq:torque_arm}
\end{equation}
The force-to-torque conversion can also be derived from virtual work. If the mechanism configuration is determined by $\theta$ and $\Delta\theta$, the closed-loop constraint of the four-bar mechanism is
\begin{equation}
f(\theta,\Delta\theta)=0 ,
\label{eq:closure}
\end{equation}
and the differential relation is
\begin{equation}
\frac{d\theta}{d\Delta\theta}
=
-\frac{\partial f/\partial \Delta\theta}{\partial f/\partial \theta}.
\label{eq:theta_delta}
\end{equation}
The negative sign in \eqref{eq:theta_delta} follows directly from the implicit differentiation of the closed-loop constraint. Under quasi-static and frictionless assumptions, the virtual work balance between the input torque $\tau_B$ and spring force $F_s$ is
\begin{equation}
\tau_B\,\mathrm{d}(\Delta\theta)
=
F_s(\theta)\,\mathrm{d}l_s .
\label{eq:virtual_work}
\end{equation}
Since $\mathrm{d}l_s=(dl_s/d\theta)\mathrm{d}\theta$, the input torque can be written as
\begin{equation}
\tau_B(\theta,\Delta\theta)
=
F_s(\theta)
\frac{d l_s}{d\theta}
\frac{d\theta}{d\Delta\theta}.
\label{eq:force_to_torque}
\end{equation}
In \eqref{eq:virtual_work} and \eqref{eq:force_to_torque}, $\tau_B$ is also treated as a signed generalized torque under the chosen positive input direction. The sign of the torque is therefore determined by the angular transmission ratio and the spring-length derivative. If the spring torque is instead defined as the restoring torque exerted by the elastic element on the input, the opposite sign convention should be used consistently.
Equations \eqref{eq:torque_arm} and \eqref{eq:force_to_torque} describe torque generation from the viewpoints of equivalent moment arm and virtual work transmission. The nonlinear output of the CFB-CTE does not require a nonlinear spring. Instead, it arises from the combined variation of spring elongation, equivalent moment arm, and angular transmission ratio.

\noindent\hspace*{1em}3) Target torque and equivalent stiffness. The equivalent output stiffness is defined as the derivative of output torque with respect to input angle:
\begin{equation}
K_B(\theta)=
\frac{\partial \tau_B(\theta)}{\partial \theta}.
\label{eq:stiffness}
\end{equation}
A quadratic target curve is used to describe the desired torque-angle relationship:
\begin{equation}
\tau_{ref}(\Delta\theta)
=
a_1\Delta\theta+a_2\Delta\theta^2 ,
\label{eq:target_torque}
\end{equation}
with the corresponding target stiffness
\begin{equation}
K_{ref}(\Delta\theta)
=
\frac{\partial \tau_{ref}}{\partial \Delta\theta}
=
a_1+2a_2\Delta\theta .
\label{eq:target_stiffness}
\end{equation}
Equations \eqref{eq:target_torque} and \eqref{eq:target_stiffness} provide the target curves for structural optimization. The aim is not to simply increase or decrease spring stiffness. Instead, the optimization searches for mechanism parameters that make $\tau_B(\theta)$ approximate $\tau_{ref}(\Delta\theta)$ within the target workspace while maintaining acceptable inertia, mass, and installation offset.

\noindent 4) Structural cost indices. The elastic unit must satisfy not only the target torque curve, but also lightweight and low-inertia requirements. If only torque error is minimized, the optimizer may select long links or heavy components, which degrades dynamic performance. Equivalent inertia, total mass, and center-of-mass offset are therefore included in the optimization model.

Let the mass and self-rotational inertia of component $i$ be $m_i$ and $I_i$, respectively. Let $\boldsymbol r_{G,i}(\theta)$ denote its center-of-mass position, and let $\boldsymbol r_O$ denote the output reference point. The equivalent structural inertia is
\begin{equation}
I_{eq}(\theta)=
\sum_i
\left[
I_i+
m_i
\left\|
\boldsymbol r_{G,i}(\theta)-\boldsymbol r_O
\right\|^2
\right].
\label{eq:ieq}
\end{equation}
At discrete sampling angles $\theta_j$, the average equivalent inertia is
\begin{equation}
\bar I_{eq}
=
\frac{1}{N}
\sum_{j=1}^{N}
I_{eq}(\theta_j).
\label{eq:mean_ieq}
\end{equation}
The total mass is
\begin{equation}
M=\sum_i m_i .
\label{eq:mass}
\end{equation}
The center-of-mass offset evaluates the motion of the global center of mass:
\begin{equation}
\bar d_G
=
\frac{1}{N}
\sum_{j=1}^{N}
\left\|
\boldsymbol r_G(\theta_j)-\boldsymbol r_{G,0}
\right\|.
\label{eq:dg}
\end{equation}
Equations \eqref{eq:ieq}--\eqref{eq:dg} extend the design problem from single-curve fitting to a multi-objective optimization that considers output behavior and engineering feasibility.

\section{Parameter Optimization}
\subsection{Objective Function}
The optimization uses three weighted objectives: torque-curve fitting, inertia-related structural cost, and structural mass. The center-of-mass offset is included as a sub-index of the inertia-related structural cost rather than as an independent weight term. The analytical model uses $l_1$--$l_4$, $\theta_0$, and $k$ as core variables. To include link dimensions and gear-related components in the numerical implementation, the design vector is extended to 11 parameters:
\begin{equation}
\boldsymbol x=[l_1,l_2,l_3,l_4,\theta_0,k,b,t,t_g,\alpha_o,\rho_h]^T ,
\label{eq:xopt}
\end{equation}
where $b$ and $t$ are the link width and thickness, $t_g$ is the gear thickness, $\alpha_o$ is the ratio between the gear outer radius and pitch radius, and $\rho_h$ is the lightening-hole radius ratio. The combined objective function is defined as a dimensionless fitness function:
\begin{equation}
J(\boldsymbol x)
=
w_TJ_T(\boldsymbol x)
+w_IJ_I(\boldsymbol x)
+w_MJ_M(\boldsymbol x),
\label{eq:objective}
\end{equation}
where $J_T$, $J_I$, and $J_M$ correspond to the torque-curve error, inertia-related structural cost, and mass cost derived from the mechanical indices defined in \eqref{eq:rmse}--\eqref{eq:mass}. Since these quantities have different physical units, each term is normalized before weighted summation. The weights $w_T$, $w_I$, and $w_M$ satisfy
\begin{equation}
w_T+w_I+w_M=1,\qquad w_i\ge 0,\; i\in\{T,I,M\}.
\label{eq:weights}
\end{equation}
The torque-curve fitting error is first evaluated by RMSE:
\begin{equation}
E_T(\boldsymbol x)
=
\sqrt{
\frac{1}{N}
\sum_{j=1}^{N}
\left[
\tau_B(\Delta\theta_j;\boldsymbol x)
-
\tau_{ref}(\Delta\theta_j)
\right]^2
}.
\label{eq:rmse}
\end{equation}
The dimensionless objective terms used in \eqref{eq:objective} are then defined as
\begin{equation}
J_T(\boldsymbol x)=
\frac{E_T(\boldsymbol x)}
{\max_j|\tau_{ref}(\Delta\theta_j)|},
\label{eq:jt_norm}
\end{equation}
\begin{equation}
J_I(\boldsymbol x)=
\frac{\bar I_{eq}(\boldsymbol x)}{I_{ref}}
+\lambda_G
\frac{d_G(\boldsymbol x)}{d_{G,ref}},
\label{eq:ji_norm}
\end{equation}
\begin{equation}
J_M(\boldsymbol x)=
\frac{M_{tot}(\boldsymbol x)}{M_{ref}}.
\label{eq:jm_norm}
\end{equation}
Here, $\bar I_{eq}$, $d_G$, and $M_{tot}$ are obtained from \eqref{eq:ieq}, \eqref{eq:dg}, and \eqref{eq:mass}, respectively. The reference scales are $I_{ref}=90000~\mathrm{g\,mm^2}$, $d_{G,ref}=45.0~\mathrm{mm}$, and $M_{ref}=100~\mathrm{g}$, and the center-of-mass offset coefficient is $\lambda_G=0.25$.

\subsection{Geometric Feasibility and Parameter Constraints}
To ensure manufacturability and assembly feasibility, the design variables are constrained by parameter bounds and geometric checks. The link-length bounds are $l_1\in[40,80]~\mathrm{mm}$, $l_2,l_3\in[70,120]~\mathrm{mm}$, and $l_4\in[40,100]~\mathrm{mm}$. The initial installation angle is $\theta_0\in[0.3,1.2]~\mathrm{rad}$, and the spring stiffness is $k\in[0.008,0.030]~\mathrm{N/mm}$. The extended variables satisfy $b\in[10,20]~\mathrm{mm}$, $t\in[3.0,5.0]~\mathrm{mm}$, $t_g\in[4.0,8.0]~\mathrm{mm}$, $\alpha_o\in[1.05,1.35]$, and $\rho_h\in[0,0.75]$.

In addition to these bounds, the four-bar mechanism must avoid unreachable configurations, obvious dead points, and link interference within the target workspace. The spring length must meet installation and elongation requirements, and the output torque curve must remain continuous. The shortest-link condition requires $l_1$ to be the shortest link and $l_1+l_4\le l_2+l_3$. The initial spring length is constrained to be close to $80~\mathrm{mm}$, with an allowable relative deviation of 5\%. A hard penalty is applied if the gear radius, lightening hole, or kinematic calculation becomes infeasible.

\begin{figure}[htbp]
\centering
\includegraphics[width=0.96\columnwidth]{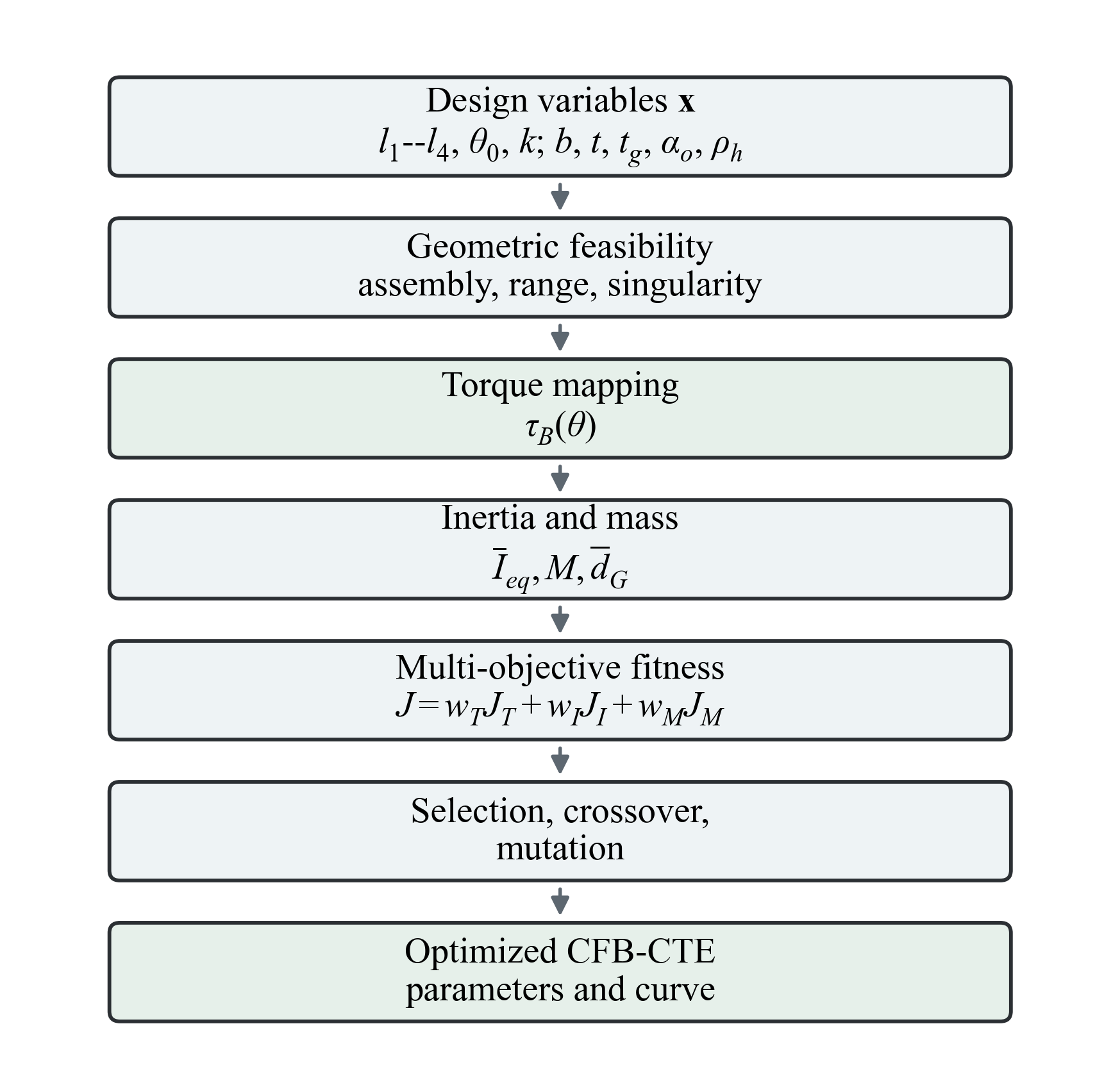}
\caption{CFB-CTE parameter optimization workflow. The procedure integrates geometric feasibility checking, torque-curve calculation, inertia and mass evaluation, and genetic algorithm operations within one optimization framework.}
\label{fig:flow}
\end{figure}

A genetic algorithm is adopted for global parameter search. Genetic algorithms are suitable for nonlinear and multimodal optimization problems with implicit geometric constraints \cite{goldberg1989}. They are therefore appropriate for the non-convex design space formed by closed-chain geometry, spring elongation, and structural constraints in the CFB-CTE. As shown in Fig.~\ref{fig:flow}, the optimization first defines the design-variable bounds and fixed material parameters. It then generates an initial population randomly. Each candidate is checked for geometric feasibility, after which the torque curve, equivalent inertia, mass, and center-of-mass offset are evaluated. Selection, crossover, and mutation are then performed until the convergence criterion or maximum generation number is reached. The population size is 260, the maximum number of generations is 3500, the stall generation limit is 600, and the function tolerance is $10^{-8}$.

\subsection{Weight Analysis and Parameter Selection}
\begin{figure}[!t]
\centering
\includegraphics[width=0.96\columnwidth]{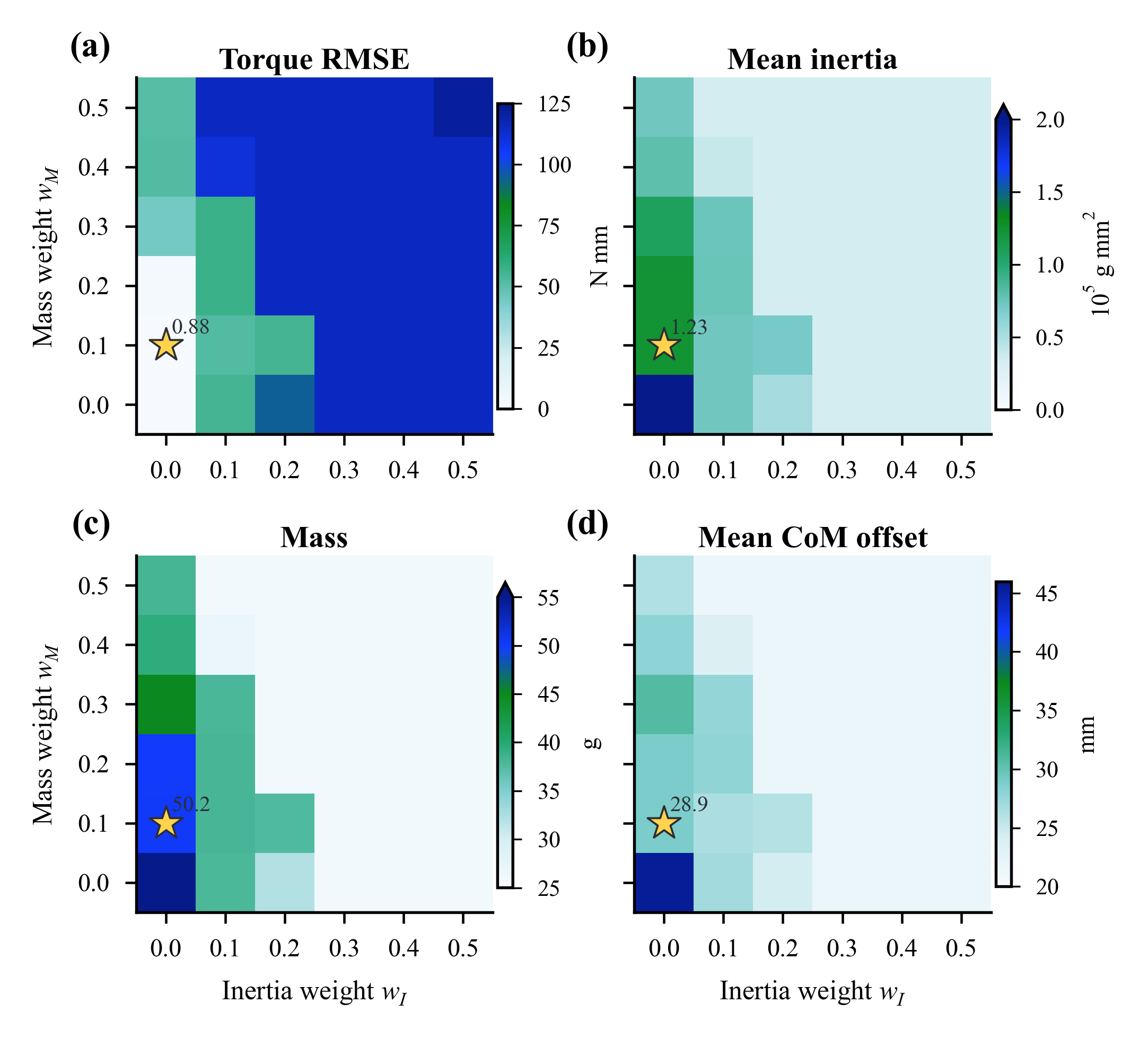}
\caption{Optimization results under different weight combinations. (a) Torque RMSE. (b) Average equivalent inertia. (c) Total mass. (d) Average center-of-mass offset. The star marks the selected compromise weight combination.}
\label{fig:heatmap}
\end{figure}

To examine the influence of different objective terms, the inertia-related and mass weights are scanned. The scanning step is 0.1, the maximum values of $w_I$ and $w_M$ are 0.5, and $w_T=1-w_I-w_M$. A total of 36 feasible weight combinations are obtained. Fig.~\ref{fig:heatmap} shows the distributions of torque RMSE, average equivalent inertia, total mass, and average center-of-mass offset in the weight plane. The minimum torque error occurs at $w_I=0,w_M=0$, with an RMSE of $0.554~\mathrm{N\,mm}$. However, this torque-only candidate increases the average equivalent inertia to $1.97\times10^6~\mathrm{g\,mm^2}$ and the mass to $345~\mathrm{g}$. Compared with the selected compromise design, these values are about 16.0 and 6.9 times larger, respectively, indicating that the dominant penalty of the torque-only solution is the excessive dynamic inertia rather than a small increase in mass. At the other extreme, the minimum-mass solution reaches $25.7~\mathrm{g}$, but its torque RMSE increases to $115~\mathrm{N\,mm}$; the minimum center-of-mass offset solution similarly gives $21.6~\mathrm{mm}$ with an RMSE of $115~\mathrm{N\,mm}$. These lightweight solutions therefore satisfy the geometric bounds but no longer preserve the required torque-shaping behavior. The final solution is selected by first retaining candidates with RMSE below $2.0~\mathrm{N\,mm}$ and then ranking them using a normalized weighted score. Under this accuracy constraint, the selected weight set is $w_T=0.9,w_I=0,w_M=0.1$, resulting in a torque RMSE of $0.883~\mathrm{N\,mm}$, an average equivalent inertia of $1.23\times10^5~\mathrm{g\,mm^2}$, a total mass of $50.2~\mathrm{g}$, and a center-of-mass offset of $28.9\pm0.4~\mathrm{mm}$ over the sampled motion range. The corresponding optimized and derived parameters used for multibody modeling are listed in Table~\ref{tab:param}.

\begin{table}[!t]
\caption{Main design parameters and optimization results of the CFB-CTE}
\label{tab:param}
\centering
\scriptsize
\setlength{\tabcolsep}{2pt}
\renewcommand{\arraystretch}{1.03}
\begin{tabularx}{\columnwidth}{@{}>{\raggedright\arraybackslash}p{0.18\columnwidth}>{\centering\arraybackslash}p{0.13\columnwidth}>{\centering\arraybackslash}p{0.11\columnwidth}X>{\centering\arraybackslash}p{0.13\columnwidth}@{}}
\toprule
Parameter & Symbol & Unit & Range/constraint & Result \\
\midrule
Torque weight & $w_T$ & -- & $w_T=1-w_I-w_M,\; w_T\ge0$ & 0.9000 \\
Inertia weight & $w_I$ & -- & $[0,0.5]$, step 0.1 & 0.0000 \\
Mass weight & $w_M$ & -- & $[0,0.5]$, step 0.1 & 0.1000 \\
Link 1 & $l_1$ & mm & $[40,80]$ & 63.487 \\
Link 2 & $l_2$ & mm & $[70,120]$ & 70.386 \\
Link 3 & $l_3$ & mm & $[70,120]$ & 70.000 \\
Link 4 & $l_4$ & mm & $[40,100]$ & 63.876 \\
Initial angle & $\theta_0$ & rad & $[0.3,1.2]$ & 1.2000 \\
Spring stiffness & $k$ & N/mm & $[0.008,0.030]$ & 0.0300 \\
Link width & $b$ & mm & $[10,20]$ & 10.000 \\
Link thickness & $t$ & mm & $[3.0,5.0]$ & 3.000 \\
Gear thickness & $t_g$ & mm & $[4.0,8.0]$ & 4.000 \\
Gear ratio factor & $\alpha_o$ & -- & $[1.05,1.35]$ & 1.0500 \\
Hole ratio & $\rho_h$ & -- & $[0,0.75]$ & 0.7500 \\
Total mass & $M_{tot}$ & g & Computed & 50.2 \\
Gear inertia & $I_g$ & $\mathrm{g\,mm^2}$ & Computed & 15870 \\
\bottomrule
\end{tabularx}
\end{table}

\begin{figure}[!t]
\centering
\includegraphics[width=0.96\columnwidth]{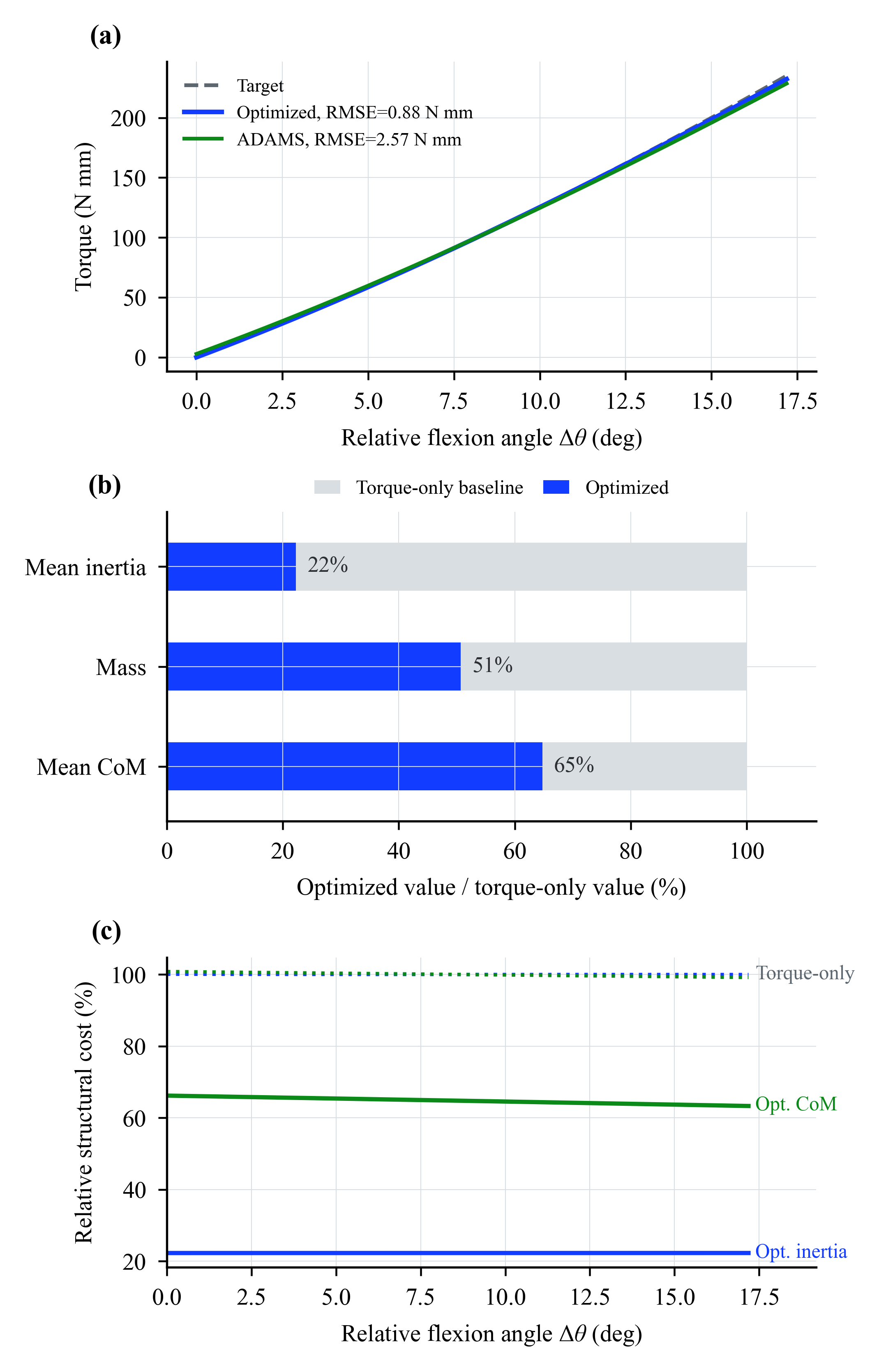}
\caption{Torque fidelity and structural cost of the selected compromise design. (a) Target torque curve, analytical optimized curve, and ADAMS post-processed curve. (b) Average structural cost of the compromise design relative to the independent torque-only optimized design. (c) Normalized equivalent inertia and center-of-mass offset over the relative flexion angle $\Delta\theta$.}
\label{fig:inertia}
\end{figure}

Fig.~\ref{fig:inertia} compares the compromise design, ADAMS post-processed curve, and independent torque-only optimized design. To avoid angular misalignment caused by different initial installation angles, the horizontal axis is defined as the relative flexion angle $\Delta\theta$. The analytical curve of the compromise design follows the target curve with an RMSE of $0.883~\mathrm{N\,mm}$. The ADAMS post-processed torque curve has an RMSE of $2.57~\mathrm{N\,mm}$. The independent torque-only optimized design provides a slightly smaller fitting error, but its average equivalent inertia is $5.52\times10^5~\mathrm{g\,mm^2}$, which is 4.49 times that of the compromise design. Its total mass is 1.97 times that of the compromise design. The center-of-mass offset decreases from $44.6~\mathrm{mm}$ to $28.9\pm0.4~\mathrm{mm}$ in the compromise design, where the value is reported as mean $\pm$ SD over 200 sampled angles.

\section{Simulation Validation}
\subsection{Multibody Model}
After parameter optimization, the optimized CFB-CTE must be verified in a multibody environment. A multibody model containing the crossed four-bar mechanism, elastic element, transmission connection, and output joint is built in Automatic Dynamic Analysis of Mechanical Systems (ADAMS). The link dimensions, spring stiffness, and initial installation angle are taken from the optimization results. The spring force acts on the output side through the mechanism geometry, forming the torque-angle relation corresponding to the analytical model.

The ADAMS model is used for independent structural validation rather than controller validation. No stiffness controller or target stiffness trajectory is introduced in this work. The reference response in Fig.~\ref{fig:structure_response} is obtained from the analytical kinematic model under the same prescribed input motions, and is used as a baseline for checking whether the optimized multibody mechanism preserves the intended input-output angular transmission.

\subsection{Comparison Between Reference and ADAMS Responses}
\begin{figure}[!t]
\centering
\includegraphics[width=0.96\columnwidth]{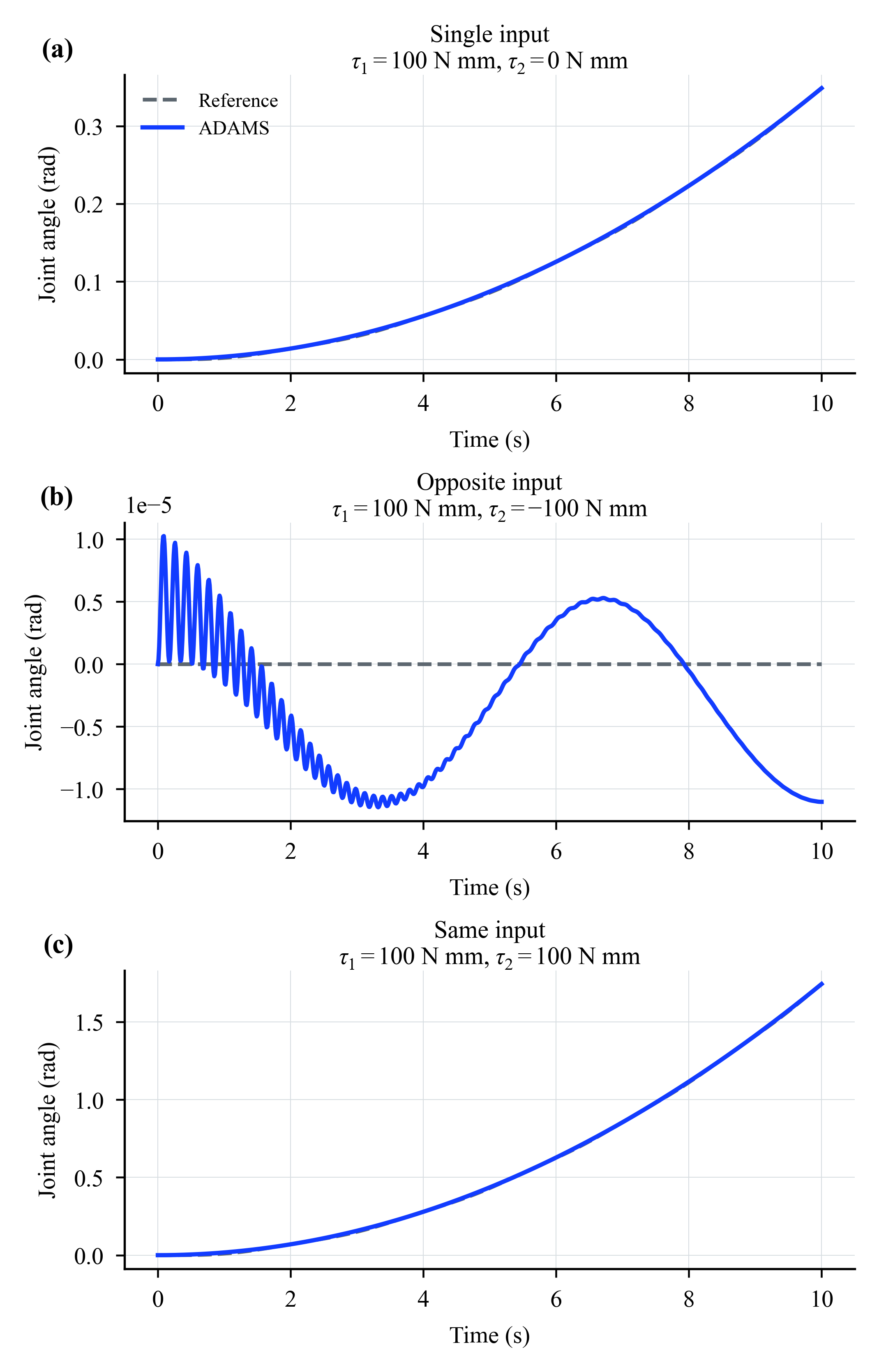}
\caption{Structural responses of the VSA under different input conditions in ADAMS. (a) Single-input condition. (b) Opposite-input condition. (c) Same-input condition. The reference curves are calculated from the analytical kinematic model under the same input motions, and the three panels compare the corresponding joint-angle responses.}
\label{fig:structure_response}
\end{figure}

Fig.~\ref{fig:structure_response} compares the ADAMS model response with the analytical reference response under representative input conditions. The reference curves are not prescribed by a control algorithm; instead, they are calculated from the optimized mechanism geometry and the same input-angle histories used in the ADAMS model. Under the single-input condition, the output joint produces continuous displacement and reaches a final joint angle of $0.3485~\mathrm{rad}$. The RMSE is $8.69\times10^{-4}~\mathrm{rad}$, and the maximum absolute error is $1.53\times10^{-3}~\mathrm{rad}$. Under the opposite-input condition, the joint angle remains near zero, with an RMSE of $6.58\times10^{-6}~\mathrm{rad}$ and a maximum absolute deviation of $1.15\times10^{-5}~\mathrm{rad}$. This condition corresponds to antagonistic preload rather than position output. Under the same-input condition, the output angle increases to $1.7427~\mathrm{rad}$. The RMSE is $3.70\times10^{-3}~\mathrm{rad}$, and the maximum absolute error is $7.57\times10^{-3}~\mathrm{rad}$. These three input cases correspond to single-side drive, antagonistic preload, and coordinated position output, respectively.

The ADAMS model includes multibody geometric connections and realistic motion constraints. The agreement between the simulated and analytical reference responses supports the kinematic consistency and engineering feasibility of the optimized geometric parameters.

\section{Discussion}
The results show that the crossed four-bar mechanism can convert a linear spring response into a nonlinear torque-angle relation by shaping the spring elongation path and equivalent moment arm. Compared with a design that only pursues torque fitting, the compromise design maintains a torque RMSE of $0.883~\mathrm{N\,mm}$ while reducing the total mass to $50.2~\mathrm{g}$. It also decreases the average equivalent inertia to $1.23\times10^5~\mathrm{g\,mm^2}$ and the center-of-mass offset to $28.9\pm0.4~\mathrm{mm}$ over the sampled motion range. These results indicate that the CFB-CTE can serve not only as a target-curve fitting mechanism, but also as a lightweight elastic branch for antagonistic VSAs.

The ADAMS validation further shows that the optimized parameters can reproduce single-input, opposite-input, and same-input responses under multibody constraints. The present study is still limited to structural design and simulation validation. Manufacturing errors, friction, assembly clearance, and spring installation deviation have not yet been considered. Future work will focus on prototype fabrication and joint-bench experiments, direct measurement of torque-angle curves and stiffness regulation range, robust optimization under manufacturing uncertainty, and integration with position-stiffness control strategies for cyclic motion, impact mitigation, and human-robot interaction tasks.

\section{Conclusion}
This paper proposed a crossed four-bar compliant transmission structure for a variable stiffness actuator. The architecture analysis, parameter optimization, and ADAMS validation were completed. The main conclusions are as follows.

1) The crossed four-bar mechanism reshapes a linear spring response into a nonlinear torque-angle relation through spring-length variation and equivalent moment-arm variation.

2) A multi-objective function combining torque error, inertia-related structural cost, and mass constrains both output behavior and structural cost.

3) The multibody joint-angle responses agree with the analytical reference responses, supporting the kinematic feasibility of the optimized CFB-CTE for VSA structural design.

The proposed method provides a geometry-shaping optimization route for elastic-unit design in VSAs and establishes a parameter basis for prototype development, robustness analysis, and experimental validation.

\section*{Acknowledgment}
This work was supported in part by the National Natural Science Foundation of China under Grant 52475291, the Beijing Natural Science Foundation under Grant L222139, the Fundamental Research Funds for the Central Universities under Grants GW2025-71 and YWF-23-Q-1031, and the 2025 Doctoral Student Special Program of the Young Elite Scientists Sponsorship Program by CAST.


\begin{thebibliography}{00}
\bibitem{desantis2008} A. De Santis, B. Siciliano, A. De Luca, and A. Bicchi, ``An atlas of physical human-robot interaction,'' \emph{Mechanism and Machine Theory}, vol. 43, pp. 253--270, 2008.
\bibitem{haddadin2009} S. Haddadin, A. Albu-Schaffer, and G. Hirzinger, ``Requirements for safe robots: Measurements, analysis and new insights,'' \emph{The International Journal of Robotics Research}, vol. 28, pp. 1507--1527, 2009.
\bibitem{bicchi2004} A. Bicchi and G. Tonietti, ``Fast and soft-arm tactics,'' \emph{IEEE Robotics \& Automation Magazine}, vol. 11, pp. 22--33, 2004.
\bibitem{pratt1995} G. A. Pratt and M. M. Williamson, ``Series elastic actuators,'' in \emph{Proc. IEEE/RSJ International Conference on Intelligent Robots and Systems}, 1995, pp. 399--406.
\bibitem{vanderborght2013} B. Vanderborght \emph{et al.}, ``Variable impedance actuators: A review,'' \emph{Robotics and Autonomous Systems}, vol. 61, pp. 1601--1614, 2013.
\bibitem{wolf2016} S. Wolf \emph{et al.}, ``Variable stiffness actuators: Review on design and components,'' \emph{IEEE/ASME Transactions on Mechatronics}, vol. 21, pp. 2418--2430, 2016.
\bibitem{grioli2015} G. Grioli \emph{et al.}, ``Variable stiffness actuators: The user's point of view,'' \emph{The International Journal of Robotics Research}, vol. 34, pp. 727--743, 2015.
\bibitem{laurin1991} K. F. Laurin-Kovitz, J. E. Colgate, and S. D. R. Carnes, ``Design of components for programmable passive impedance,'' in \emph{Proc. IEEE International Conference on Robotics and Automation}, 1991, pp. 1476--1481.
\bibitem{english1999} C. English and D. Russell, ``Implementation of variable joint stiffness through antagonistic actuation using rolamite springs,'' \emph{Mechanism and Machine Theory}, vol. 34, pp. 27--40, 1999.
\bibitem{tonietti2005} G. Tonietti, R. Schiavi, and A. Bicchi, ``Design and control of a variable stiffness actuator for safe and fast physical human/robot interaction,'' in \emph{Proc. IEEE International Conference on Robotics and Automation}, 2005, pp. 526--531.
\bibitem{vanham2007} R. Van Ham, B. Vanderborght, M. Van Damme, B. Verrelst, and D. Lefeber, ``MACCEPA, the mechanically adjustable compliance and controllable equilibrium position actuator,'' \emph{Robotics and Autonomous Systems}, vol. 55, pp. 761--768, 2007.
\bibitem{jafari2010} A. Jafari, N. G. Tsagarakis, B. Vanderborght, and D. G. Caldwell, ``A novel actuator with adjustable stiffness (AwAS),'' in \emph{Proc. IEEE/RSJ International Conference on Intelligent Robots and Systems}, 2010, pp. 4201--4206.
\bibitem{tsagarakis2011} N. G. Tsagarakis, I. Sardellitti, and D. G. Caldwell, ``A new variable stiffness actuator (CompAct-VSA): Design and modelling,'' in \emph{Proc. IEEE/RSJ International Conference on Intelligent Robots and Systems}, 2011, pp. 378--383.
\bibitem{jafari2011} A. Jafari, N. G. Tsagarakis, and D. G. Caldwell, ``AwAS-II: A new actuator with adjustable stiffness based on adaptable pivot point and variable lever ratio,'' in \emph{Proc. IEEE International Conference on Robotics and Automation}, 2011, pp. 4638--4643.
\bibitem{groothuis2014} S. S. Groothuis, G. Rusticelli, A. Zucchelli, S. Stramigioli, and R. Carloni, ``The variable stiffness actuator vsaUT-II: Mechanical design, modeling, and identification,'' \emph{IEEE/ASME Transactions on Mechatronics}, vol. 19, pp. 589--597, 2014.
\bibitem{petit2010} F. Petit, M. Chalon, W. Friedl, M. Grebenstein, A. Albu-Schaffer, and G. Hirzinger, ``Bidirectional antagonistic variable stiffness actuation: Analysis, design and implementation,'' in \emph{Proc. IEEE International Conference on Robotics and Automation}, 2010, pp. 4189--4196.
\bibitem{albu2010} A. Albu-Schaffer \emph{et al.}, ``Dynamic modelling and control of variable stiffness actuators,'' in \emph{Proc. IEEE International Conference on Robotics and Automation}, 2010, pp. 2155--2162.
\bibitem{hussain2018} I. Hussain, A. Albalasie, M. I. Awad, L. Seneviratne, and D. Gan, ``Modeling, control, and numerical simulations of a binary-controlled variable stiffness actuator,'' \emph{Frontiers in Robotics and AI}, vol. 5, article 68, 2018.
\bibitem{wang2025} C. Wang, Z. Zhang, Y. Xiao, P. Gao, and X. Liu, ``Dynamic modelling and experimental investigation of an active-passive variable stiffness actuator,'' \emph{Actuators}, vol. 14, article 169, 2025.
\bibitem{mommersteeg1996} T. J. A. Mommersteeg, L. Blankevoort, R. Huiskes, J. G. M. Kooloos, and J. M. G. Kauer, ``Characterization of the mechanical behavior of human knee ligaments: A numerical-experimental approach,'' \emph{Journal of Biomechanics}, vol. 29, no. 2, pp. 151--160, 1996.
\bibitem{li2004} G. Li, L. E. DeFrate, S. Zayontz, E. Most, J. F. Suggs, and H. E. Rubash, ``In situ forces of the anterior and posterior cruciate ligaments in high knee flexion: An in vitro investigation,'' \emph{Journal of Orthopaedic Research}, vol. 22, no. 2, pp. 293--297, 2004.
\bibitem{carlin1996} G. J. Carlin, G. A. Livesay, C. D. Harner, Y. Ishibashi, H. S. Kim, and S. L.-Y. Woo, ``In-situ forces in the human posterior cruciate ligament in response to posterior tibial loading,'' \emph{Annals of Biomedical Engineering}, vol. 24, no. 2, pp. 193--197, 1996.
\bibitem{fox1998} R. J. Fox, C. D. Harner, M. Sakane, G. J. Carlin, and S. L.-Y. Woo, ``Determination of the in situ forces in the human posterior cruciate ligament using robotic technology: A cadaveric study,'' \emph{The American Journal of Sports Medicine}, vol. 26, no. 3, pp. 395--401, 1998.
\bibitem{goldberg1989} D. E. Goldberg, \emph{Genetic Algorithms in Search, Optimization, and Machine Learning}. Reading, MA, USA: Addison-Wesley, 1989.
\end{thebibliography}
\end{document}